\documentclass[twocolumn,showpacs,preprintnumbers,amsmath,amssymb,aps,prb,superscriptaddress]{revtex4-2}

\pdfoutput=1
\usepackage[pdftex]{graphicx}

\usepackage[english]{babel}
\usepackage{hyperref}
\usepackage{amsmath}
\usepackage{float}
\usepackage{soul}

\usepackage{amssymb}
\usepackage[utf8]{inputenc}
\usepackage[T1]{fontenc}
\usepackage{graphicx}
\usepackage{braket}
\usepackage{xcolor}

\usepackage[caption=false]{subfig}

\usepackage{cancel}
\usepackage{color,soul}

\usepackage{epstopdf}
\usepackage{enumitem,kantlipsum}
\usepackage{physics}
\usepackage{bbold}
\usepackage{array}

\usepackage[T1]{fontenc}
\usepackage[sc]{mathpazo}
\makeatletter

\usepackage[caption=false]{subfig}
\usepackage{latexsym}
\usepackage{braket}

\@ifundefined{showcaptionsetup}{}{%
	\PassOptionsToPackage{caption=false}{subfig}}
\usepackage{subfig}
\makeatother

\begin{document}
\title{High-root topological edge-state bands}

\author{R. G. Dias}
\email{rdias@ua.pt}
\affiliation{Department of Physics $\&$ i3N, University of Aveiro, 3810-193 Aveiro,
Portugal}

\author{L. Madail}
\affiliation{Department of Physics $\&$ i3N, University of Aveiro, 3810-193 Aveiro,
Portugal}
\affiliation{International Iberian Nanotechnology Laboratory, 4715-310 Braga, Portugal}

\author{A. M. Marques}
\affiliation{Department of Physics $\&$ i3N, University of Aveiro, 3810-193 Aveiro,
Portugal}

\date{\today}

\begin{abstract}
This paper presents a complex band analysis of one-dimensional (1D) square and high-root topological insulators (HRTIs). We show that edge-state bands of HRTIs are sliced sections of impurity bands of a uniform tight-binding chain. A simplified topological characterization of HRTIs with generalized boundary conditions is carried out based on the existence of edge-state bands in the infinite  HRTI and the restrictions imposed by the boundary conditions. Edge states in finite or semi-infinite 1D HRTIs are shown to be a subset of evanescent states of the infinite system and mapped onto impurity states of the uniform chain with effective energy-dependent edge potentials. The latter result allows the determination of the edge state levels without needing the diagonalization of real space or bulk Hamiltonians.
\end{abstract}
\maketitle

\section{Introduction}
\label{sec:intro}

Square-root topological insulators (SRTIs) are systems whose Hamiltonians are square roots  of a conventional topological insulator (TI) Hamiltonian \cite{Ezawa2020, Arkinstall2017,Pelegri2019a,Kremer2020}, which one designates as the parent Hamiltonian. These systems exhibit  in-gap edge states at non-central energy gaps whose origin can be traced back to the existence of  edge eigenstates at the central gap of the parent Hamiltonian. High-root topological insulators (HRTIs) generalize this definition either by recursively applying square-root operations \cite{Marques2021,Dias2021,Marques2021a,Marques2023} or a single N-root operation to a known TI (the latter usually generates non-Hermitian models)  \cite{Viedma2024}. 
 
High-root topological insulators exhibit spectra that are fractal-like versions of simple topological insulator (TI) spectra  (obtained by repeatedly squaring the Hamiltonian) and a multitude of finite energy edge states \cite{Marques2021,Dias2021,Marques2021a,Marques2023}. They represent an interesting case where the effect of edge local potentials is not apparent, especially concerning the finite energy edge states.  
HRTIs are a generalization of square-root TIs  \cite{Arkinstall2017,Pelegri2019a} and, similarly to them, they can be experimentally realized in artificial platforms such as acoustic lattices \cite{Yan2020,Cheng2022,Wu2023,Geng2024}, photonic lattices \cite{Cui2023,Kang2023,Yan2021,Wei2024}, electrical circuits \cite{Song2020,Song2022}, Floquet systems \cite{Bomantara2022,Zhou2022,Cheng2022}, and more.

 While the tenfold way classifies TIs based on global symmetries (time-reversal, particle-hole, chiral) and dimensionality, in contrast, HRTIs are defined by their algebraic relationship to a parent Hamiltonian, and this does not require new symmetry classes. Instead, HRTIs leverage existing symmetries (e.g., chiral symmetry) in the parent TI Hamiltonian to protect edge states at non-zero energies in the gaps corresponding to the parent TI.  This means that these edge states at non-zero energies are protected against perturbations  that do not break the chiral symmetry of the parent Hamiltonian \cite{Viedma2024}. 

The topological characterization of HRTIs is not straightforward since the usual expressions for topological invariants often lead to non-quantized values in these systems \cite{Kremer2020,Marques2019,Springborg2004,Kudin2007,Miert2017,Rhim2017,Lin2018,Arkinstall2017,Kremer2020a,Midya2018,Zhang2019,Pelegri2019,Pelegri2019b}.
In fact, square-root TIs were first described as TIs where their relation to the parent TI could be used to overcome this problem \cite{Kremer2020a,Pelegri2019,Ke2020,Yoshida2021,Wu2021,Ding2021,Matsumoto2023}.
Furthermore, the usual characterization of 1D topological insulators relies on topological invariants extracted from the bulk behavior of the system, where bulk is interpreted as the inner region of a finite system \cite{Asboth2016, Hasan2010}. Bulk properties are expected to have the behavior observed in the infinite system. This assumption should, however, be considered carefully. For example, let us consider a tight-binding ring with $N$ sites. As $N$ grows, the momenta $k=2 \pi (j/N)$ with $j=0,\cdots, N-1$ densify the interval $[0,2\pi[$, but clearly $k/(2 \pi)$ will always be a rational number. In contrast, all real values  $k/(2 \pi)$ are present in the set of solutions of the eigenvalue equation of the infinite chain. So, one should not confuse thermodynamic limit $N \rightarrow \infty$ with $N= \infty$ \footnote{A historical parallel can be drawn here with the Casimir effect \cite{Casimir1948}, which relates the attractive force between parallel metallic plates in free space to vacuum fluctuations of the electromagnetic field. There is an infinite and continuous set of vacuum modes present in free space, that reflect the fact that all real values for the frequencies are allowed. On the other hand, when the parallel plates are introduced, the vacuum modes in the confined region form an infinite yet discrete set along their normal direction, since confinement forces the quantization of frequencies to rational values only. It is ultimately the vacuum fluctuations induced by changing the set of its modes from continuous to discrete that generate the attractive force.
Analogously, the theoretical solution for the blackbody radiation spectrum obtained by Planck famously required him to assume the infinite set of energies for each allowed frequency to be discrete rather than continuous.}.
Additionally, the limit can produce different results depending on the starting point. For example, starting from the topological phase of the Su-Schrieffer-Heeger (SSH) chain \cite{Su1979} with open boundary conditions, the spectrum when the number of unit cells  $N_{uc} \rightarrow \infty$ will have two zero energy states. In contrast, no gapless levels are present starting from the trivial phase. 
\begin{figure*}[t]
\includegraphics[width=1\textwidth]{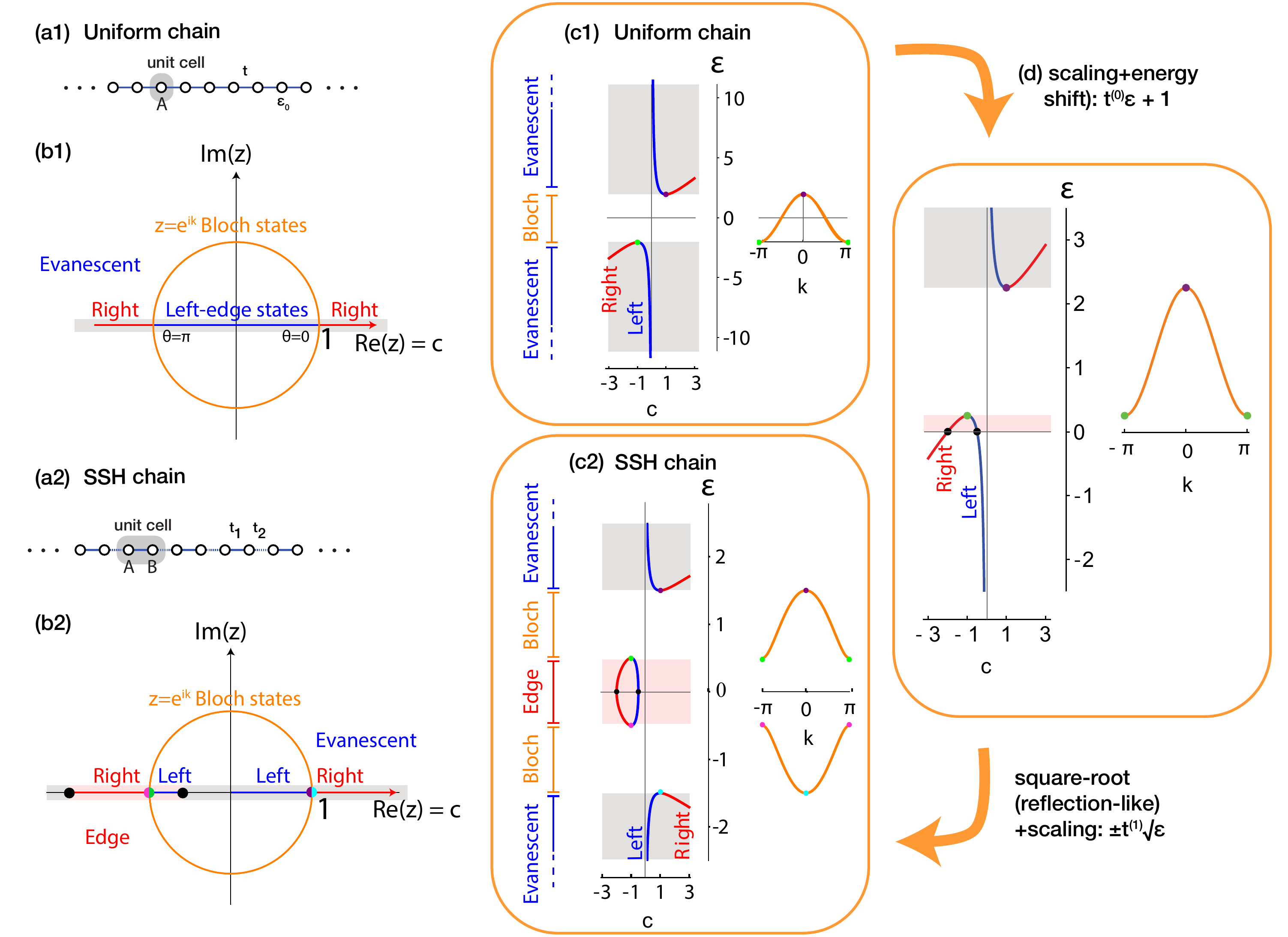}
\caption{(a1,b1,c1) Infinite uniform tight binding chain: (a1) lattice; (b1) real energy solutions of the tight-binding eigenvalue equation in the complex $z$-plane which are  Bloch (orange circle) and edge states  (blue/red line); (c1) Band structure including the Bloch band (orange curve) and evanescent state bands (blue and red curves, corresponding to left and right decaying states, respectively). (a2,b2,c2) The same as above for the infinite SSH chain with $t_1=1/2 $ and $t_2=1$. 
(c1,c2,d) Illustration of how an impurity band of an infinite uniform chain generates the edge-state band in the SSH chain in the square-root operation. In (c1) and (c2), the evanescent state and Bloch bands share the states at the inversion invariant momenta. The same-colored circles indicate these shared points. (d) The band structure obtained by scaling and shifting the uniform chain spectrum, with a part of the bottom impurity band crossing the zero energy level (pink region). Two extra black dots are present in (d), indicating the intersection points of the bottom impurity band with the $\varepsilon=0$ level, and the corresponding points are also shown in (c2). The values of $t^{(0)}$ and $t^{(1)}$ in the square-root process are respectively $4/5$ and $(5/4)^{1/2}$ (see App.~D for the definition of these parameters).
\label{fig:1}}
\end{figure*}
The latter implies that the SSH model (and 1D HRTIs) is sometimes regarded as a poor example of a topological insulator due to the ambiguity in defining the bulk topological invariants in the infinite SSH chain \cite{Cayssol2021}. That is, the choice of the unit cell determines whether Zak's phase is trivial or non-trivial, but in the infinite system, that choice should be physically irrelevant.

The Squarable sine-cosine chain is a simple example of a 1D  HRTI \cite{Dias2021}.
Squarable sine-cosine chains (SSC(n)) are 1D tight-binding models built upon the SSH model, generalized to 2n sites per unit cell. The recursive squaring (and energy shifts) of the SSC(n) Hamiltonian generates a sequence of self-similar Hamiltonians with progressively smaller unit cells. Each squaring step preserves the structural form of the model while renormalizing hopping terms via sine-cosine parameterizations, enabling a "Matryoshka" hierarchy of topological phases. Furthermore, this process introduces distinct chiral symmetries at each level, protecting edge states in multiple energy gaps and creating a nested structure of topological robustness. This has described in detail in Ref. \cite{Dias2021}.

In this paper, motivated by the above discussion, we adopt a simplified and pragmatic topological characterization of square- and high-root 1D topological insulators \cite{Dias2021,Marques2021, Ezawa2020} with generalized boundary conditions (which imply the usual topological invariants are insufficient to characterize the behavior of these systems). The generalized boundary conditions consist of coupling to clusters, which, as we will show, lead to energy-dependent boundary conditions of the TI chain when a decimation scheme is applied \cite{Mukherjee2020,Bandy2021,Mukherjee2022,Mukherjee2024,Biswas2025}, whereby the cluster is sequentially contracted to a single site.
We note that a similar type of decimation scheme, consisting of a renormalization method called isospectral reduction \cite{Smith2019}, has been recently applied to uncover models with latent symmetries that enable the generation of flat bands \cite{Morfonios2021}, non-Hermitian skin effect \cite{Eek2024} and latent topological states \cite{Rontgen2023,Eek2024b,Eek2024c}.
Our topological characterization of 1D HRTIs involves two steps:
(i) an analysis of the infinite system using complex band theory (see \cite{Reuter2016} and References there in) that generalizes the usual analysis of square and higher-root topological insulators by including edge state bands. In this step, we show that edge state bands of HRTIs are sliced parts of impurity bands of a uniform chain;
(ii) the topological characterization of the finite system using the number of edge states in the multiple gaps of the HRTIs as topological invariant. 

Note that our results imply that the edge-state bands observed in infinite HRTI chains can be probed by introducing clusters at the boundaries of finite or semi-infinite HRTI chains. Closed-form equations for the determination of these cluster-induced edge-state levels are obtained that do not require diagonalization of real space or bulk Hamiltonians and that reveal that these states  can be mapped onto impurity Tamm states within uniform chains with energy-dependent potentials. These cluster-induced  edge states of HRTIs are a subset of the infinite system edge-state bands and in the case of the ${\cal SSC}(n)$  chain, can be interpreted as the $n$-root of impurity states of the uniform chain with effective edge potentials, with positive integer $n$. 

The paper is organized as follows.
In Sec.~II, we address edge state bands in the infinite SSH chain and extend this analysis to the case of infinite Squarable Sine-Cosine chains (which are HRTIs). 
Sec.~III discusses a method for determining edge state energies in finite or semi-infinite 1D HRTIs. 
Sec.~IV  proposes an energy-dependent bipartition of 1D lattices with edge potentials.
In Sec.~V,  we apply the methods of the previous sections to particular cases of 1D TIs.
Finally, in Sec.~VI we conclude.

\section{Infinite chains}
As stated above, we present a simplified topological characterization of 1D topological insulators, with two layers of topological characterization, one corresponding to the infinite system and the other due to the restrictions imposed by the boundary conditions. 

Complex band theory \cite{Reuter2016} tells us the real eigenspectrum of an infinite tight-binding system is continuous if complex momentum is allowed, with \emph{non-orthogonal and non-normalizable}  evanescent states bands completing the \emph{orthogonal but non-normalizable}  Bloch state bands spectrum, that is,  filling in the energy gaps as shown in Fig.~1 in the cases of the uniform and SSH chains. We adopt the designation of edge states for evanescent states lying between  Bloch bands (note that these states may disappear at a gap-closing point while no gap-closing is possible for the other evanescent states we designate as impurity states).
All bands are determined by diagonalizing the infinite system Hamiltonian. More precisely, we assume the ansatz $|\psi(z)\rangle= \ket{z}\otimes|u(z)\rangle$ with $\langle j |z\rangle= z^j $and solve for $h(z)\ket{u(z)}=\epsilon(z)\ket{u(z)},$ where $h(z)$ is the Hamiltonian matrix of infinite dimension, imposing the condition of the real $\epsilon(z)$. 

In Fig.~1(b1) and (b2), we show that two types of $z$-solutions of the eigenvalue equation appear in the case of the infinite uniform and  SSH chains (see App.~A and C for more details): (i) a phase $z=e^{i k}$, where $k$ is the (real) momentum, and the respective eigenstates are Bloch states; (ii) a real $z$ which we label as $c$,  and the respective eigenstates are evanescent states.
Recalling that the SSH chain is the square root of the uniform chain, it is simple to relate both models' evanescent-state bands and eigenstates. As shown in Fig.~1, the edge-state band of the infinite SSH chain is generated in the square-root operation by (i) the shift of the bands of the uniform chain so that the bottom impurity band crosses the zero energy level [see Fig.~1(d)], and (ii) an additional reflection-like step of the positive part of the spectrum [pink and grey regions in Fig.~1(d) in what concerns the evanescent states] due to the square-root operation, that generates the loop-like energy dispersion curve of the edge-state band of the SSH chain.

Underlying the behavior described above for the appearance of the edge-state bands is the usual relation between square-root TIs and the parent Hamiltonian \cite{Ezawa2020}, that is, the square of the infinite Hamiltonian matrix of the bipartite SRTI,
\begin{equation}
\sqrt{H}=\begin{bmatrix}
0 & H^{AB}\\
H^{BA} & 0\\\end{bmatrix},
\end{equation}
generates a two-block diagonal matrix (the blocks have the infinite dimension of the respective sublattices, $A$ and $B$), with one of the blocks corresponding to the parent Hamiltonian (and the other being the residual block) with an energy shift due to the uniform diagonal terms of the parent block,
\begin{equation}
H={\sqrt{H}}^2=\begin{bmatrix}
H_1&0\\
0&H_2\\
\end{bmatrix}=H_1\oplus H_2,
\end{equation}
where $H_1=H^{AB} H^{BA}$ and $H_2=H^{BA} H^{AB}$.
Eigenstates of the SRTI Hamiltonian remain eigenstates of the squared Hamiltonian, independently of being Bloch or evanescent states, $\sqrt{H} \vert \psi \rangle = \varepsilon \vert \psi \rangle \Rightarrow H \vert \psi \rangle = \varepsilon^2 \vert \psi \rangle$. We also have $ H_1 \vert \psi_A \rangle = \varepsilon^2 \vert \psi_A \rangle$ and $ H_2 \vert \psi_B \rangle = \varepsilon^2 \vert \psi_B \rangle$ where $\vert \psi_{A/B} \rangle$ are the components of the eigenstate $\vert \psi\rangle$ associated to  sublattices $A$ and $B$.
\begin{figure}[tb]
	\includegraphics[width=0.95\columnwidth]{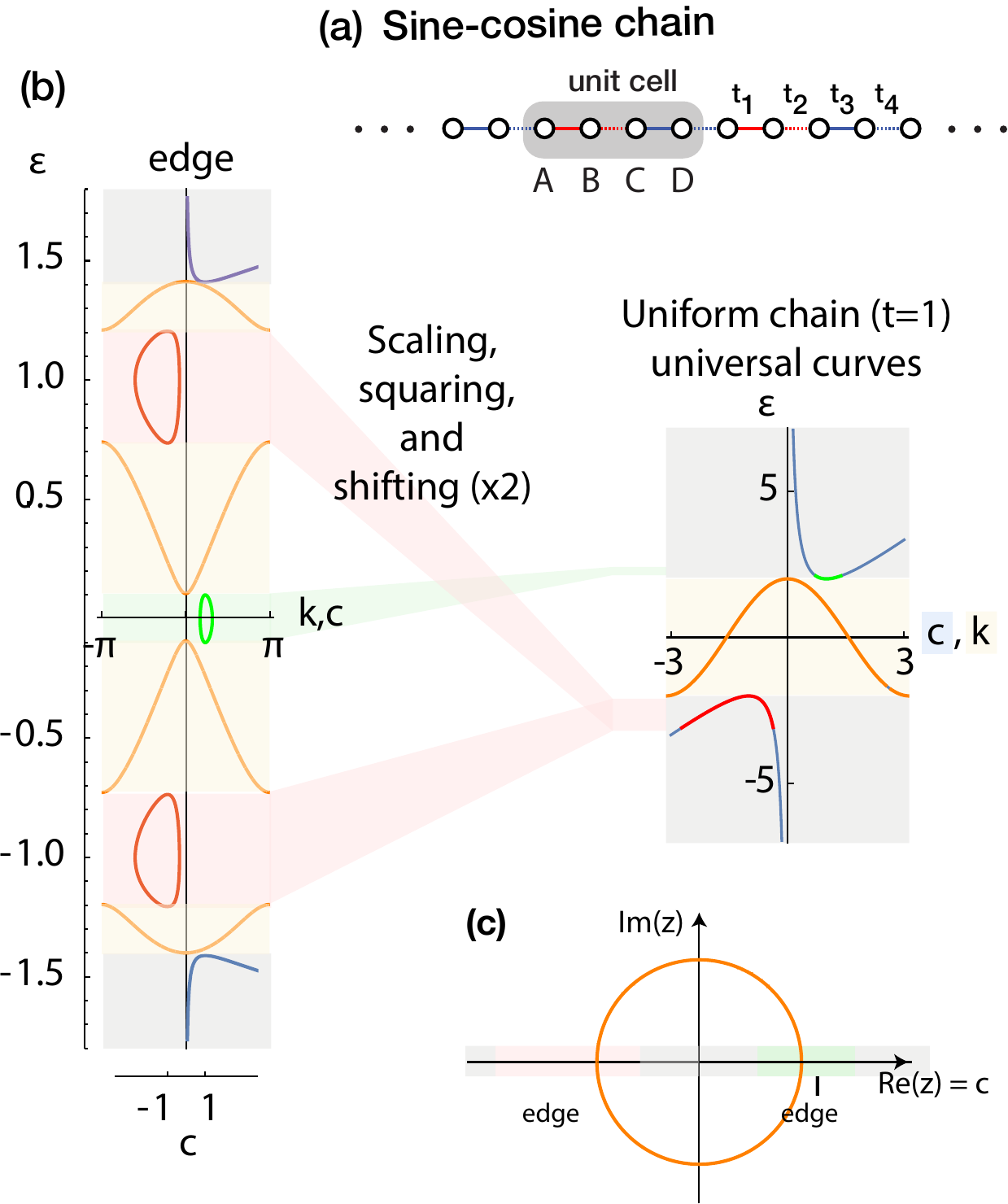}
	\caption{The ${\cal SSC}(2)$ chain [shown in (a)] can be topologically characterized by the mapping of the edge state bands onto the uniform chain dispersion curves, in the squaring process shown in (b). Parameters of the ${\cal SSC}(2)$ chain: $\theta_1=1.094$, $\theta_2=0.618$, leading to $\{t^{(0)}, t^{(1)}\} = \{0.771, 0.324  \}$. (c) Solutions with real eigenenergies in the $z$-plane with color shading indicating the respective regions of the ${\cal SSC}(2)$ band structure.}
    \label{fig:SSC2}
\end{figure}

As mentioned in the Introduction, the usual topological classification of the infinite SSH chain implies a nonphysical particular choice of the infinite chain unit cell.
A more basic topological invariant (independent of this unit cell choice but with less information about the bulk behavior than the winding vector or the Zak's phase) is to interpret the opening of the SSH band gap as the 1D analog of adding a hole to a sphere generating a toric surface. Therefore, the number of band gaps is the 1D equivalent of the genus number.
A simple example of multiple gaps is provided by the high roots of the uniform chain, described in App.~D, which are designated as Squarable Sine- Cosine chains, ${\cal SSC}(n)$ \cite{Dias2021}. In Fig.~\ref{fig:SSC2}(a), we depict a segment of an infinite ${\cal SSC}(2)$ chain with four sites in the unit cell and hopping terms $\{t_1,t_2,t_3,t_4\}=\{\sin(\theta_1),\cos(\theta_1),\sin(\theta_2),\cos(\theta_2)\}$, with $\theta_1=1.094$ and $\theta_2=0.618$ [or equivalently $\{t^{(0)}, t^{(1)}\} = \{0.771, 0.324 \}$ in the Matryoshka squaring process, see App.~D].
The corresponding band structure is shown in the left panel of Fig.~\ref{fig:SSC2}(b). 
By squaring this model twice, one maps the edge state bands onto the impurity bands of the uniform chain, as shown in Fig.~\ref{fig:SSC2}(b). This mapping is another example of a topological invariant that can be defined in infinite chains and relies on non-orthogonal edge state bands (or, equivalently, energy gaps). 
Only the top and bottom edge state bands of the HRTI map onto the fourth quadrant (negative energy and negative $c$) of the uniform chain since impurity bands after the first root are always on the positive $c$ axis, see Fig.~\ref{fig:SSC2}(b).

\begin{figure}[b]
\includegraphics[width=0.9\columnwidth]{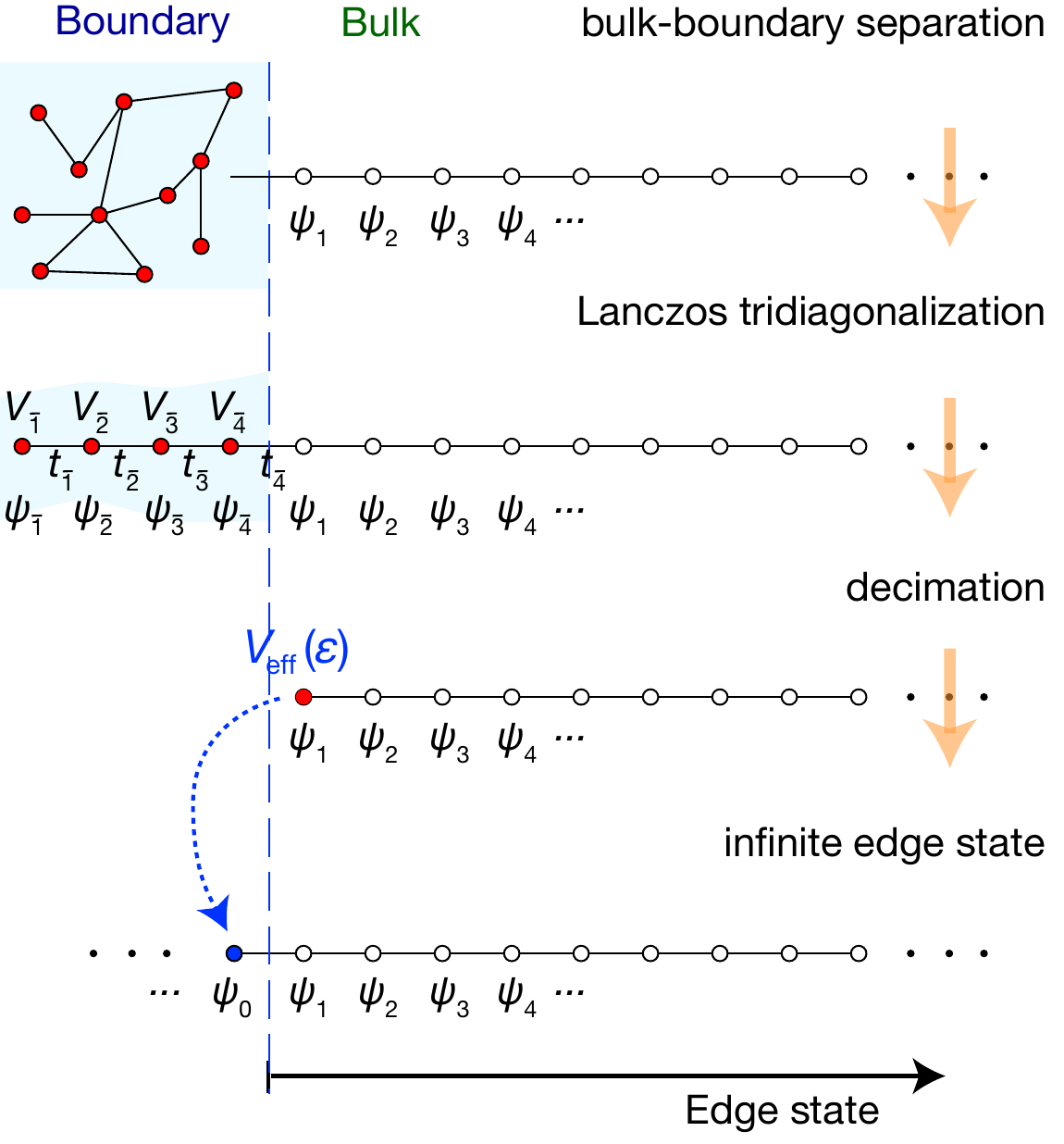}
\caption{Steps justifying the main argument of this paper, namely, that the tight-binding eigenstates in the bulk are truncated versions of the eigenstates of the infinite chain: (i) separation of the semi-infinite chain in a boundary region and a bulk region; (ii) Lanczos tridiagonalization of the boundary; (iii) decimation leading to an effective single-site local potential; (iv) selection of a particular edge state of the in-gap continuum set of edge states of the infinite chain.
 }
 \label{fig:stepsdiagram}
\end{figure}
\section{Semi-infinite and finite chain and local edge potentials}
We now discuss the second layer of topological characterization in the context of semi-infinite or finite topological insulators, which arises from the restrictions imposed by boundary conditions. This discussion builds on a central argument: the presence of boundary conditions (with or without local edge potentials) selects specific solutions of the eigenvalue equation for the infinite tight-binding chain. In the case of open boundary conditions (OBC), these solutions can be either Bloch or edge-like states. In finite systems with OBC, the degeneracy of Bloch states with opposite momenta is used to find eigenstates that satisfy both boundary conditions. These typically lead to harmonic states, which have zero amplitudes at the virtual sites at the edges of the finite chain. Similarly, the boundary conditions select specific edge states from the non-orthogonal edge state band of the infinite chain, potentially combining them. Note that the usual topological characterization becomes a special case where the topological invariants, determined by the boundary conditions, match those of the infinite system with a particular choice of unit cell.

Furthermore, edge state bands can be probed repeatedly by introducing energy-dependent boundary conditions. These conditions expand the Hilbert space (that is, there are more eigenstates of the Hamiltonian than sites) and, in the case of a chain, can be interpreted as the boundary effect of an additional cluster of sites at the chain edge. This cluster effectively introduces an energy-dependent local potential at the edge, as shown in Fig.~3, and the respective eigenstates are truncated versions of eigenstates of the infinite chain (in the corresponding region) with amplitudes in the boundary region that compensate precisely for the absence of the rest of the infinite chain. 

A more detailed explanation of the previous arguments involves four steps: (i) separation of the semi-infinite chain in a boundary region and a bulk region; (ii) Lanczos tridiagonalization of the boundary; (iii) decimation leading to an effective single-site local potential; (iv) selection of a particular edge state of the in-gap continuum set of edge states of the infinite chain. These steps are similar to those in the recursive Green's function method applied to electronic transport calculations (see e.g. \cite{Lewenkopf2013}). Furthermore, steps (2) and (3) can be replaced by alternative approaches that also reduce the boundary cluster to an effective single-site local potential (see App.~E for one example). Still, our approach is numerically efficient by avoiding the diagonalization of the cluster.

(i) Bulk-boundary separation: a central argument in this paper is the separation of the semi-infinite chain (or ribbon) in a boundary region and a bulk region. Here, we identify the bulk region as the region with eigenstate amplitudes that match those of an eigenstate of the infinite chain; that is, the tight-binding eigenstates in the bulk are truncated versions of eigenstates of the infinite chain (in the corresponding region). 

(ii) Lanczos tridiagonalization of the boundary: the Lanczos algorithm reduces a general matrix to a tridiagonal matrix adopting a new basis, the Lanczos vectors, where the only unchanged basis state compared with the original basis is the first state. Therefore, applying the Lanczos method to the set of Wannier states of the cluster plus site 1 of the bulk (starting from the Wannier state $\vert 1 \rangle$), we reduce this set to a linear chain with different hopping terms $\{ t_{\bar{n}} \}$ and local potentials $\{ V_{\bar{n}} \}$ along the chain, stopping at the bulk site $j=1$ where the local potential is zero. The integer $\bar{n}$ indexes the basis states of the tridiagonal Hamiltonian matrix associated with the cluster, as shown in Fig.~3. 
Note that the number of finite $\{ t_{\bar{n}} \}$ in this process can be, in particular cases, less than the number of cluster sites (see Sec.~V.A for an example), reflecting the decoupling of part of the cluster states from the chain. This decoupled subspace will generate levels independent of the chain parameters, and the respective states are entirely localized in the cluster. 

(iii) Decimation leading to an effective single-site local potential: the new boundary region obtained in the previous step can be reduced to an energy-dependent local potential at the rightmost site of the boundary region (third diagram of Fig.~\ref{fig:stepsdiagram}) by Gaussian elimination of the amplitudes in the boundary region starting from the leftmost site and using the eigenvalue equations $\epsilon(c) \psi_{\bar{n}}= V_{\text{eff}}(c)^{(\bar{n})} \psi_{\bar{n}} + t_{\bar{n}} \psi_{\overline{n+1}} $ until we are left with the effective potential at the site $j=1$ of the bulk region. The Gaussian elimination sequence leads to  recurrence relations (note the difference in indexing in amplitudes and potentials)
\begin{equation}
\left\{
\begin{array}{l}
\displaystyle
V_{\text{eff}}(c)^{(\bar{1})}  = V_{\bar{1}},
\\
\displaystyle
V_{\text{eff}}(c)^{(\bar{n})} = V_{\bar{n}} + \dfrac{t_{\overline{n-1}}^2}{\epsilon(c) - V_{\text{eff}}(c)^{(\overline{n-1})}  }, \quad \bar{n} \neq 1.
\end{array}
\right.
\end{equation}
Note that for each $c$, one may have several values of $\varepsilon(c)$ as shown in Fig.~2, and solutions must be found for each of these branches of $\varepsilon(c)$.

These recurrence equations assume that the eigenvalues satisfy $\epsilon(c) - V_{\text{eff}}(c)^{(\overline{n-1})}\neq 0$. If instead equality occurs for some particular $\bar{j}$, $\epsilon(c) - V_{\text{eff}}(c)^{(\bar{j})}=0$,  the tight-binding equation above  implies $\psi_{\overline{j+1}}=0$ and we should apply the decimation starting from the site $\overline{j+2}$ to obtain this solution. Note that the full eigenstate includes the finite amplitudes at the left of the node.

(iv) Selection of an infinite edge state: the edge states are determined using the approach explained in  App.~B for the case of $V_{\text{eff}}$  independent of $c$, that is, from the relation $V_{\text{eff}} (c) \psi_{1}= t_{0\rightarrow 1}\psi_{0}$ (where $t_{0\rightarrow 1}$ is the hopping parameter between sites 0 and 1 of the infinite chain and $\psi_{0}$ is the amplitude at site 0 of the infinite chain, see next section), which will have a maximum of $N_\text{b}$ solutions,   $N_\text{b}$  being the number of sites in the boundary region. That is, the boundary region must generate the equivalent of the amplitudes on the infinite chain's first unit cell absent from the semi-infinite chain's bulk region (see last step in Fig.~3). 

Note that the orthogonality of edge states in the finite and semi-infinite chain is recovered when the amplitudes in the original boundary region are considered. Therefore, the lack of orthogonality of the edge states in the infinite chain can be ignored, assuming that the orthogonality will be recovered by introducing boundary regions in semi-infinite or finite systems.

We now make an additional comment concerning the last step. Local perturbations do not change the energy of Bloch states in an infinite chain, and the respective amplitudes are changed only locally. This energy independence reflects the fact that Bloch states are not normalizable. However, an edge state in a semi-infinite chain is normalizable, and one could expect that a local perturbation at the edge should be able to change its energy and form.
However, one has a band of edge-like states, and it is possible to find self-consistently an energy such that the perturbed region replaces the half of the infinite chain necessary to complete the semi-infinite chain, and the edge state retains its form apart from the perturbed region. That is, the perturbation selects a particular edge state of the evanescent state band. 

\section{Energy-dependent bipartition in the presence of edge potentials }
In this section, we describe how to map edge states of Squarable Sine-Cosine chains onto Tamm impurity states of the uniform chain \cite{Tamm1933,Tamm1932}. In Sec.~II, we showed that we needed to square $n$ times the initial Hamiltonian to map the edge state bands of the infinite Squarable Sine-Cosine chains onto the impurity bands of the uniform chain. We have shown in the previous section that, in the case of cluster-like boundary conditions, the complex boundary reduces to an effective energy-dependent edge potential. This edge potential makes the lattice non-bipartite, and the squaring approach described in Sec.~II does not lead to decoupled subchains. Even in the case of pristine open boundaries, the first squaring operation generates an edge potential at the subchain with the first site (different from the global energy shift of the subchain due to its lower coordination number \cite{Marques2021}). Therefore, this subchain is not bipartite.

\begin{figure}[t]
	\centering
    	\includegraphics[width=0.95\linewidth]{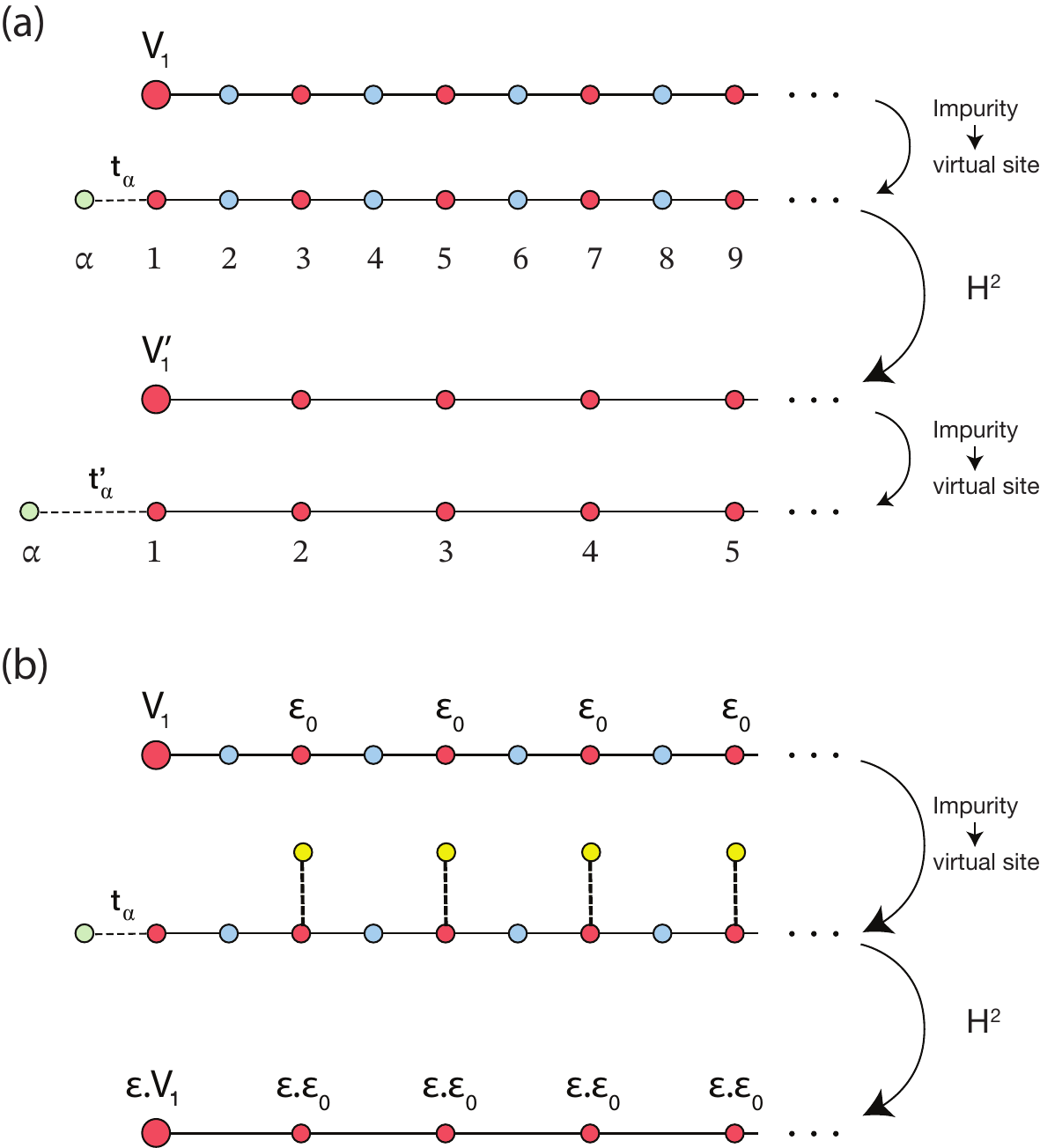}
	\caption{Onsite potentials can be replaced by energy-dependent hopping terms to virtual sites, recovering the bipartite property at each step of the $n$-squaring process. In (a), we illustrate two steps of this substitution, focusing on the subchain with the original edge site in the $n$-squaring process. In (b), the substitution is applied for modulated onsite potentials. The contribution of the virtual sites to the onsite energies is shown in the bottom subchain (other hopping terms give additional contributions). }
	\label{fig:virtualsite}
\end{figure}

Here, we avoid this problem by replacing the edge onsite potential with an energy-dependent hopping term to a virtual site, recovering the bipartite property at each step of the $n$-squaring process \cite{Pinho2022}.
Let us consider a dangling site in a chain with \emph{N} sites, as shown in Fig.~\ref{fig:virtualsite}(a).
Solving the tight-binding eigenvalue equation, we find the conditions 
$
\varepsilon \psi_1=t_{2 \rightarrow 1}\psi_2+t_{\alpha \rightarrow 1}\psi_{\alpha} $ and 
$
\varepsilon \psi_{\alpha}=t_{1 \rightarrow \alpha}\psi_{1},
$, 
where $\psi_{\alpha}$ is the amplitude at the virtual site.
We now compare this equation to that of the chain with an edge potential
$
\varepsilon \psi_1=t_{2 \rightarrow 1}\psi_2+V_1(\varepsilon)\psi_{1}
$
with an energy-dependent (in general)  local potential $V_1 (\varepsilon)$.
If $V_1 (\varepsilon)=\frac{t_{1 \rightarrow \alpha} t_{\alpha \rightarrow 1} }{\varepsilon}$, the two equations are equivalent, 
that is, the eigenstates of the chain with the dangling site also satisfy the eigenvalue equation of the chain with an energy-dependent local potential shown in Fig.~\ref{fig:virtualsite}(a). 
Importantly, the chain with a dangling site is bipartite, even if equivalent to a non-bipartite chain with a local potential. Since the local potential  $V_1 (\varepsilon)$ only depends on the product $t_{1 \rightarrow \alpha} t_{\alpha \rightarrow 1}$, 
we assume in the following  $t_{1 \rightarrow \alpha}= t_{\alpha \rightarrow 1}=t_{\alpha}=\sqrt{V_1 (\varepsilon)\epsilon}$.

We note that adding the dangling site increases the dimensionality of our Hilbert space. Therefore, one of the chain states with the dangling state must have no correspondence to the original chain. This state has zero energy and is protected by the index theorem, that is, it is a consequence of the sublattice imbalance created by the presence of the extra dangling site. 

Note additionally that, if $\varepsilon V_1 (\varepsilon)<0 $, then $t_{1 \rightarrow \alpha}= t_{\alpha \rightarrow 1}=i\sqrt{\vert V_1 (\varepsilon)\epsilon\vert}$, such that the Hamiltonian with the virtual site becomes pseudo-Hermitian (the virtual site adds a zero-energy state so that the full spectrum remains real). Nevertheless, the squared Hamiltonian is always Hermitian, and the hopping to the virtual site adds an onsite potential $V_1 \varepsilon$ to the squared Hamiltonian at the edge site. Also, if one considers the simplest case of a single site with a local potential that becomes a two-site cluster by adding the virtual site, one can easily prove that the energy has the same sign as the local potential. 

The generalization of the previous approach to multiple onsite potentials is straightforward. In Fig.~\ref{fig:virtualsite}(b), we illustrate this application in the case of modulated onsite potentials. This case will be relevant in the next section.

\section{Application to 1D TIs and HRTIs}
As stated in Sec.~\ref{sec:intro}, this paper aims to discuss the effect of local edge perturbations on finite energy edge states of HRTIs. In this section, we first discuss the simplest cases of chains with edge potentials, the semi-infinite uniform chain and the semi-infinite SSH chain, and show that the edge states in the SSH chain can be reinterpreted as square roots of (Tamm) edge impurity states of the uniform chain. We then address the case of Squarable Sine-Cosine topological insulators.

\subsection{The semi-infinite uniform tight-binding chain}
Assuming a unitary hopping parameter, the energy dispersion of edge-like states in the infinite uniform chain is $\varepsilon(c)=c+1/c$, and the absolute value  $c$ determines whether one has a left edge-like state ($\vert c\vert<1$) or a right one ($\vert c\vert>1$).

The energy of the impurity state at the edge (site 1) of the semi-infinite chain is determined using the argument that the local potential $r$ at site 1 replaces the half of the infinite chain necessary to complete the semi-infinite chain so that the edge state retains its form apart from the perturbed region. That is, at the edge of the semi-infinite chain, we have $\varepsilon(c)\psi_{1} =r\psi_{1} +\psi_{2} $ and we should have for the infinite chain  $\varepsilon(c)\psi_{1}^{\text{inf}}=\psi_{0}^{\text{inf}}+\psi_{2}^{\text{inf}}.$
Note that in this simple case, all the amplitudes are the same as those of the infinite chain normalized within the corresponding region. This is a direct consequence of assuming that the form of the edge state is the same in the non-perturbed region as that of the infinite chain. One might think that $\psi_{1}\neq\psi_{1}^{\text{inf}}$ but $\psi_{0}$ is also determined by the unperturbed tight-binding eigenvalue equation for the $\psi_{1}^{\text{inf}}$ amplitude, so one must have $\psi_{1}=\psi_{1}^{\text{inf}}$. This implies $r\psi_{1}^{\text{inf}}=\psi_{0}^{\text{inf}}.$ But we know that $ \psi_{1}^{\text{inf}}=c\psi_{0}^{\text{inf}},$ so $c=1/r.$ This implies that the energy of the impurity state is $\varepsilon(1/r)=\varepsilon(r)$, and, for the (right) impurity state to be present, one must have $\vert r\vert>1$ so that $\vert c\vert<1$.

\subsection{The semi-infinite SSH chain}
The effect of local potentials in simple 1D topological insulators (TIs) behavior is well understood. In the particular case of the SSH chain,  a local potential at the edges can change the topological phase from trivial to non-trivial or vice-versa, reflecting the creation or destruction of in-gap edge states \cite{DiLiberto2016,Marques2017a}. This fact and the ambiguity in the definition of the bulk unit cell  (and consequently in the topological invariants) have led some authors to dismiss the SSH chain as a poor example of a TI.

Squaring the semi-infinite SSH chain Hamiltonian, one obtains two decoupled uniform chains corresponding to sublattices A and B. This decoupling and the existence of a local potential in one of the subchains imply the existence of the edge states in the topological phase of the SSH chain. In fact, the zero energy edge state of the semi-infinite SSH chain is mapped onto an impurity state of subchain A with an edge local potential $-t_{2}^{2}$, if we subtract an energy $t_{1}^{2}+t_{2}^{2}$ from subchain A in order for the subchain A to have zero local potentials throughout the subchain \cite{Marques2021}. A closer mapping to the previous discussion of the semi-infinite uniform chain is possible by defining $t=t_{1}t_{2}$ and using $t$ as the energy unit. In this case,  subchain $A$ will have unitary hopping terms and a local edge potential $r=-t_2/t_1$. Recalling that a left impurity state requires $\vert r\vert>1$, we obtain the usual condition for a left edge state in the SSH chain, $\vert t_1/t_2 \vert <1$.

Let us now consider a finite local potential $V_1$ (energy-dependent in the general case) at the left edge site of the semi-infinite SSH chain. The respective Hamiltonian is not bipartite, and upon squaring,  the edge potential will generate a link between the decoupled subchains obtained in the absence of the edge potential. Nevertheless, the reasoning applied to the uniform chain can again be applied. The comparison of the edge site tight-binding equation
$
\varepsilon(c) \psi_{1A} = t_1 \psi_{1B} + V_1 \psi_{1A}
$
with the one of the infinite chain
$
\varepsilon (c) \psi_{1A} = t_1 \psi_{1B} + t_2 \psi_{0B}
$
leads to
$
V_1  \psi_{1A} = t_2 \frac{1}{c} \psi_{1B}.
$
The exact form of the edge states of the infinite SSH chain (see App.~B) gives the relation between $\psi_{1A} $ and $\psi_{1B}$, and one obtains
$$
V_1 = t_2 \dfrac{\varepsilon(c)}{c t_1 + t_2}.
$$
If one has a cluster at the left boundary, $V_1 \rightarrow V_{\text{eff}}^{(\bar{n})}(c)$ as discussed before. 

\begin{figure*}[t]
	\begin{centering}
\includegraphics[width=1\textwidth]{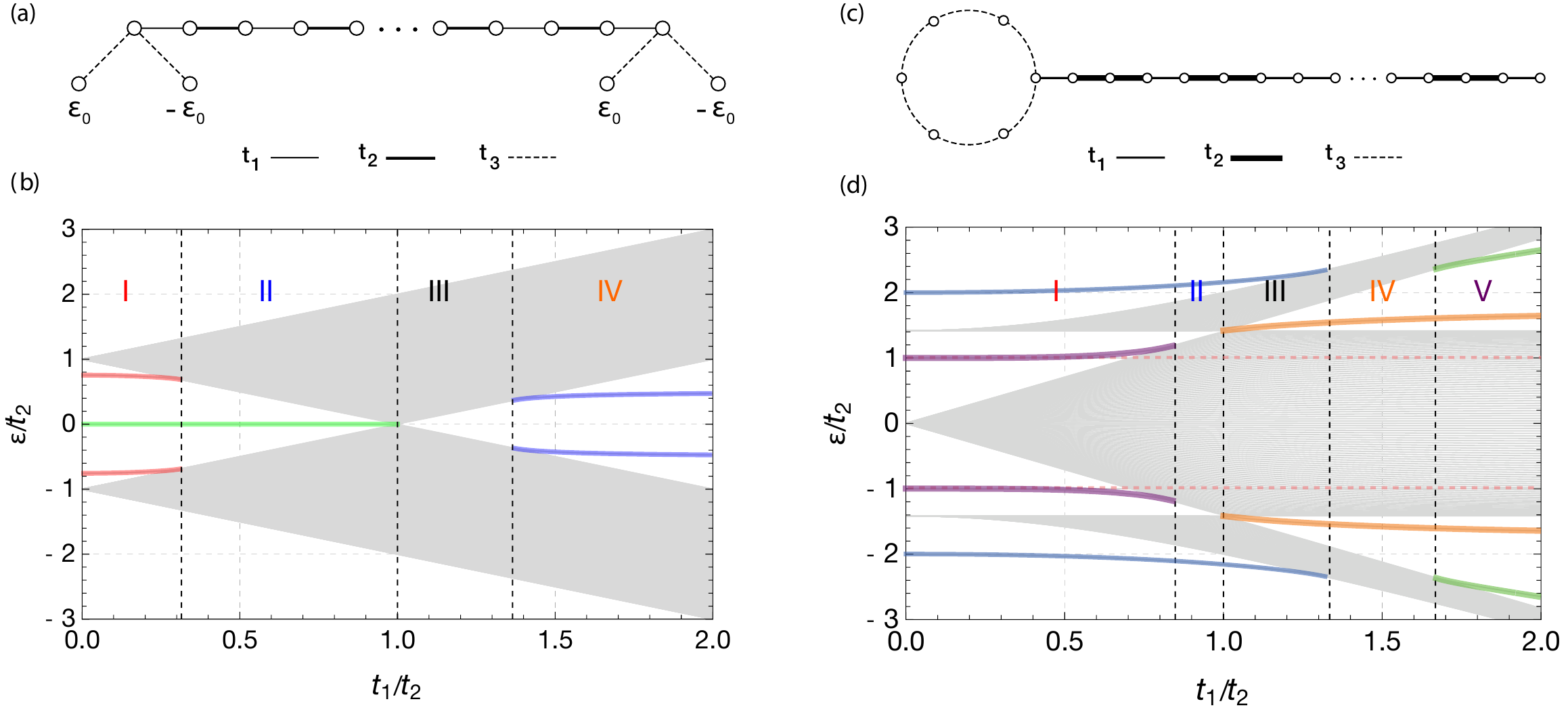}
\par\end{centering}
\caption{(a) SSH chain with two dangling sites at the left and right ends, with symmetric onsite energies $\pm \varepsilon_0$ and hopping parameter $t_3$ to the chain. (b) Respective spectrum for $\varepsilon_0=t_2/2$ and hopping parameter $t_3 = 0.4 t_2$. The black dashed lines delimit the different topological regions, with regions I and II within the infinite chain's $t_1<t_2$ phase and regions III and IV belonging to the infinite chain's $t_1>t_2$ phase. (c) $t_1 t_2 t_2 t_1$ chain connected to a six-site ring at the left edge. (d) Respective spectrum for $t_3 = t_2$. Black dashed lines delimit regions with different numbers of edge states.
\label{fig:cluster}}
\end{figure*}

Let us now address the cases shown in Fig.~\ref{fig:cluster} to illustrate the method described in Sec.~III. Fig.~\ref{fig:cluster}(a)  shows an  SSH chain with two dangling sites at the left and right ends, with symmetric onsite energies $\pm \varepsilon_0$ and hopping parameter $t_3$ to the chain.
The spectrum for $\varepsilon_0=t_2/2$ and hopping parameter $t_3 = 0.4 t_2$, shown in Fig.~\ref{fig:cluster}(b), resembles the usual one of the SSH chain, except for the presence of extra edge states that suggest the division of the usual binary topological classification (trivial/non-trivial) of the SSH chain into four regions (regions I and II within the $t_1<t_2$ part and the regions III and IV belonging to the $t_1>t_2$ part).

\begin{figure}[t]
	\begin{centering}
\includegraphics[width=1\columnwidth]{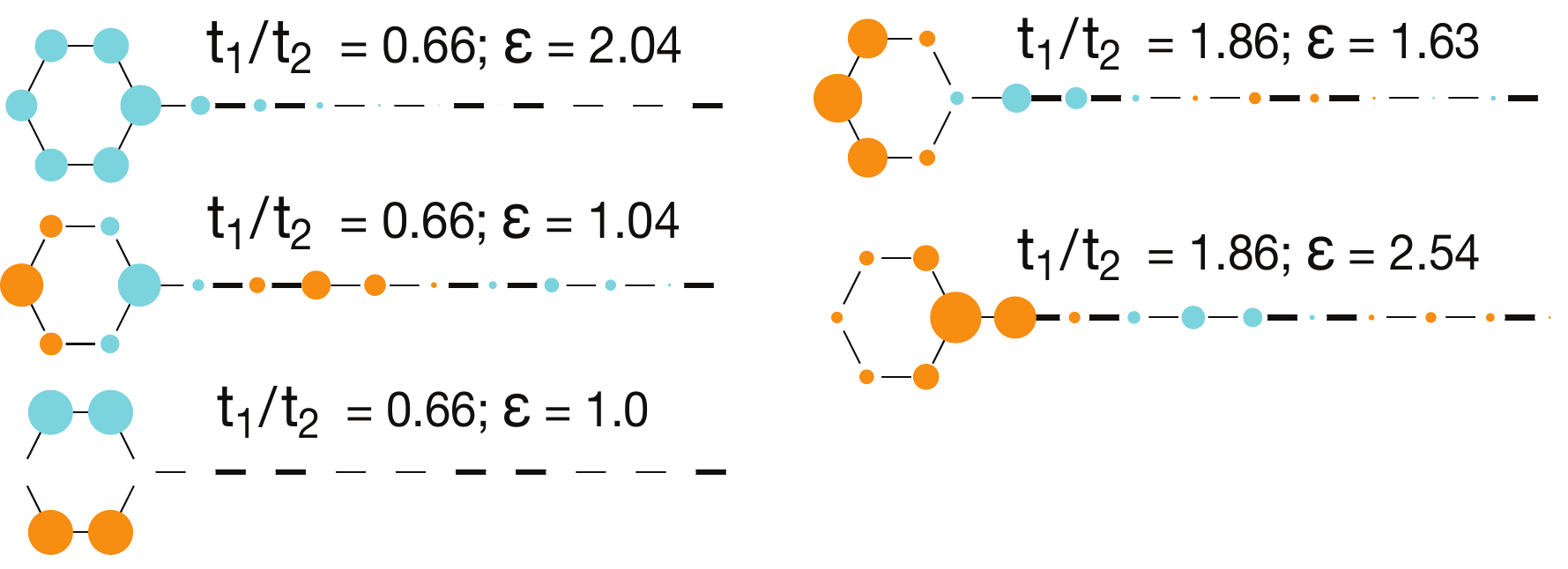}
\par\end{centering}
\caption{(a) Examples of edge states amplitude profiles of the  $t_1 t_2 t_2 t_1$ chain connected to a 6-site ring at the left edge for the top three edge-state curves in region I (profiles at the left side) of Fig.~5 and for the top two of region V (profiles at the right). The amplitudes in the $t_1 t_2 t_2 t_1$ chain are exactly those of an evanescent eigenstate of the infinite $t_1 t_2 t_2 t_1$ chain (apart from a global factor). Orange and blue indicate positive and negative amplitudes.
The profile with energy $\varepsilon=1$ has zero amplitudes in the chain, reflecting its decoupling from it. This state has energy independent of the $t_1/t_2$ ratio and generates the top red dashed line in Fig.~5(d).
\label{fig:profiles}}
\end{figure}

The spectrum in Fig.~\ref{fig:cluster}(b) was obtained by diagonalizing the full Hamiltonian matrix for a large chain. Still, the edge-state levels can be easily reproduced following the method of Sec.~II. Let us describe this application in detail for the left edge (the application to the right edge is identical). The bulk is the SSH chain, and the two dangling sites constitute the left boundary. Lanczos tridiagonalization applied to the dangling sites plus the first site of the SSH chain generates a three-site linear cluster with hopping terms $\{ t_{\bar{1} \rightarrow \bar{2} }, t_{\bar{2} \rightarrow 1 }\}=\{\varepsilon_0, \sqrt{2} t_3 \}$ and $\{ V_{\bar{1}}, V_{\bar{2}}, V_{\bar{3}} \}= \{0,0,0\}$ (where $\bar{3}=1$). Decimation leads to an effective potential 
\begin{equation}
V^{(\bar{3})}_{\text{eff}}(c)= \dfrac{(\sqrt{2} t_3)^2}{\varepsilon(c) - \dfrac{\varepsilon_0^2}{\varepsilon(c)} }.
\end{equation}
at the first site of the SSH chain.
The solutions of $V^{(\bar{3})}_{\text{eff}}(c)\psi_{1A}= t_{0\rightarrow 1}\psi_{0B}$ with $\vert c \vert <1$ determine the edge state levels.
Using the form of the edge states in the infinite SSH chain (see App.~B), this equation becomes
\begin{equation}
V_{\text{eff}}(c)^{(\bar{3})} (c) = t_{0\rightarrow 1}\dfrac{\psi_{0B}}{\psi_{1A}}=t_2 \dfrac{\varepsilon(c)}{c t_1 +t_2}.
\end{equation}
The extra edge state levels in Fig.\ref{fig:cluster} are replicated using this equation. The limits of regions I and III are easily obtained since they correspond to $c=\pm 1$, the sign depending on what is the momentum (0 or $\pi$) of the Bloch band state that touches the top or bottom of the edge state band where edge state curve ends. That is, the transition between regions corresponds as usual to the intersection of the edge state curve with the top/bottom of the Bloch bands, leading in this case to the additional transition points 
$$
\dfrac{t_1}{t_2}= 1-\left(\dfrac{t_3}{t_2}\right)^2 \pm \sqrt{\left(\dfrac{\varepsilon_0}{t_2}\right)^2
  +\left(\dfrac{t_3}{t_2}\right)^4}
$$ 
between phases I and II, and between III and IV, which were obtained replacing $c=-1$ in (5), where $V_{\text{eff}}(c)^{(\bar{3})} $ is given by (4), and $\varepsilon^2(c)$ is given by (C3) in App.~C (with $z=-1$).
As commented in Sec.~III, the zero energy edge state level is not among the solutions since $\varepsilon(c)-V_{\bar{1}}=0$ for $\varepsilon(c)=0$. In this case, one applies the method starting from site 3, which is the SSH chain's first site; therefore, one has the usual zero-energy edge state solution of the SSH eigenvalue equation. For arbitrary values of the onsite potentials of the dangling sites, the central edge state level has finite energy, and it is obtained together with the additional edge state levels using the previous approach.

For the infinite chain, the knowledge of the exact form of the eigenstates allowed us to find a polynomial equation for edge state levels in the discussion above. If the unit cell of the chain contains more than four sites, this may no longer be possible. In this case, the approach described in Sec.~IV becomes useful. We illustrate its application again in the case of the SSH chain with dangling sites at its edges shown in Fig.~\ref{fig:cluster}(a). After the decimation step, one has an effective potential $V_{\text{eff}}(c)^{(\bar{3})}$ at the first site of the SSH chain. Replacing this effective potential with the virtual site $\alpha$, one has the relation
$
V_{\text{eff}}(c)^{(\bar{3})}=\frac{t_{1 \rightarrow \alpha} t_{\alpha \rightarrow 1} }{\varepsilon}
$. Recall that this step may introduce a spurious edge state solution of zero energy due to sublattice imbalance.
Squaring the SSH chain with the virtual site, the subchain with the first site of the SSH chain has a hopping parameter $t_1 t_2$ and onsite energy $t_1^2+t_2^2$ except at the first site, where it is $t_{1 \rightarrow \alpha} t_{\alpha \rightarrow 1}+t_1^2$.
Subtracting the uniform onsite energy  $t_1^2+t_2^2$ and normalizing the hopping term, one obtains the case discussed in Sec.~V.A with $r=(t_{1 \rightarrow \alpha} t_{\alpha \rightarrow 1}-t_2^2)/(t_1 t_2)$. Recalling that $r=1/c$, one obtains
\begin{equation}
  c=  \dfrac{t_1 t_2}{V^{(\bar{3})}_{\text{eff}}(c)\varepsilon(c)-t_2^2}
  \label{eq:cformula}
\end{equation}
where the energy in the SSH chain $\varepsilon(c)$ is related to that of the uniform chain with a unitary hopping parameter, $\varepsilon_{\text{unif}}(c)$, by
$
\varepsilon^2(c)= t_1 t_2 \varepsilon_{\text{unif}}(c)+t_1^2+t_2^2
$
where
$\varepsilon_{\text{unif}}(c)=(c+1/c)$.
Plotting $\varepsilon(c)$ as a function of $c$ for $\vert c \vert <1$, an exact agreement with the left edge state curves shown in Fig.~\ref{fig:cluster}(b) is found (these curves are degenerate due to the solutions for right edge states).

\begin{figure*}[t]
	\centering
    	\includegraphics[width=1\textwidth]{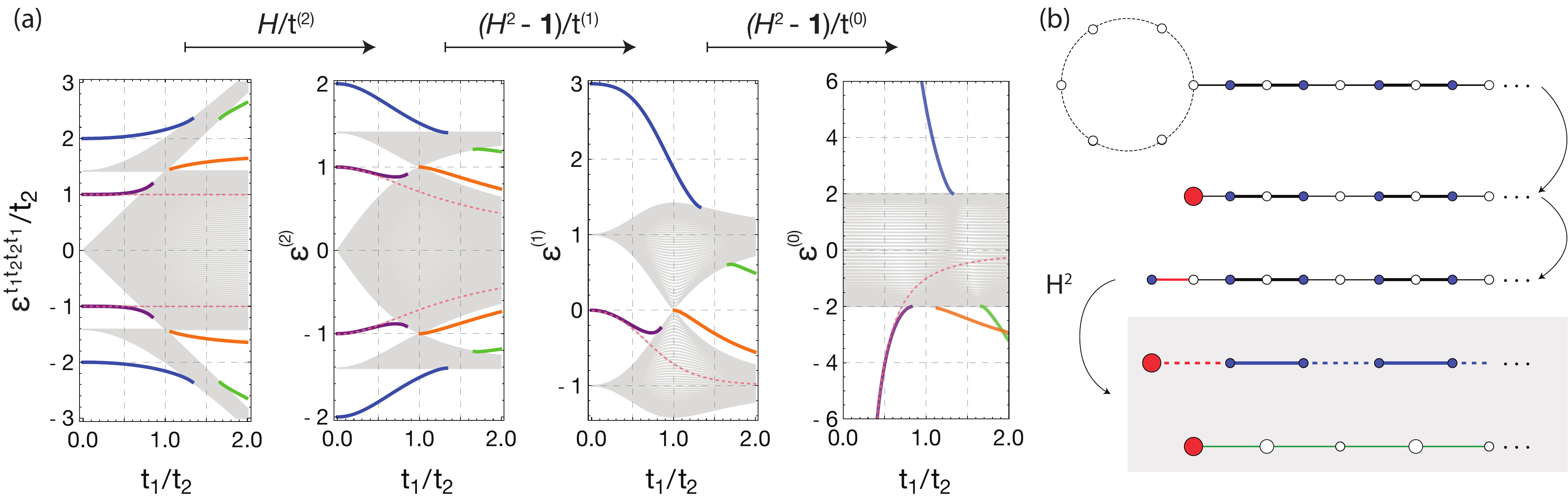}
	\caption{(a) The same spectrum as in Fig.~\ref{fig:cluster}(d) (in gray) of the $t_1 t_2 t_2 t_1$ chain (connected to a six-site ring at the left edge) in units of $t_2$, and corresponding $\varepsilon^{(2)}(c)$, 
        $\varepsilon^{(1)}(c)$, and $ \varepsilon^{(0)}(c)$,  obtained using Eqs.~(10-12). Colored curves were obtained with the method of Sec.~V.C and overlap precisely with the respective gray numerical curves. The dashed pink curves are the energy curves of the decoupled sites of the Lanczos linear cluster, which, in this case, lead to $\varepsilon=\pm t_3$ levels in the $t_1 t_2 t_2 t_1$ chain spectrum. (b) Schematics of the initial steps of the method of Sec.~V.C. In red, one has local potentials and hopping terms that are energy-dependent. Different site radius reflects different onsite energies.  }
	\label{fig:squaringOBC}
\end{figure*}

\subsection{Sine-Cosine chains}
Squarable Sine-Cosine chains, ${\cal SSC}(n)$, are simple 1D examples of HRTIs \cite{Dias2021}. When squared, these models generate two decoupled subchains: the lower-order Squarable Sine-Cosine chain, ${\cal SSC}(n-1)$, and the residual subchain. The subchains have the same spectrum in the case of periodic boundary conditions (PBC), and the relation of their energy spectrum to the original chain is
$
(\varepsilon^{(n)})^2(c)= {\cal R}^{(n-1)} \varepsilon^{(n-1)}(c)+{\cal S}^{(n-1)},
$
where ${\cal R}^{(n-1)}$ and ${\cal S}^{(n-1)}$ are respectively the energy renormalization and the energy shift in the squaring step,  ${\cal R}^{(n-1)}= t^{(n-1)}$ and ${\cal S}^{(n-1)}=1$, where $ t^{(n-1)}$ is defined in App.~D.

In the case of OBC, the spectra of the ${\cal SSC}(n-1)$ subchain and the residual subchain may differ, with the difference reflecting the existence of an additional zero-energy edge state (possibly more than one if an energy-dependent edge potential is present) in one of the subchains. Nevertheless, the multiple squaring leading to the spectrum of the uniform chain possible for PBC can also be carried out in the case of OBC,  applying the methods of Sec.~IV when needed.  

Let us illustrate this with the case of a $t_1 t_2 t_2 t_1$ chain connected to a ring of six sites with hopping parameter $t_3$, shown in Fig.~\ref{fig:cluster}(c). 
The respective spectrum, when $t_3=t_2$, is shown in Fig.~\ref{fig:cluster}(d). This spectrum displays five regions with different numbers of edge and impurity levels. Fig.~\ref{fig:profiles} shows examples of edge states amplitude profiles for the top three edge-state curves in region I (left amplitude profiles) of Fig.~\ref{fig:cluster}(d) and for the top two of region V (right amplitude profiles). 
The amplitudes in the $t_1 t_2 t_2 t_1$ chain precisely correspond to those of an evanescent eigenstate in the infinite $t_1 t_2 t_2 t_1$ chain, differing only by a global factor. Positive and negative amplitudes are represented by the colors orange and blue, respectively. The profile with $\varepsilon=1$ shows zero amplitudes in the chain, indicating its decoupling from the chain, and therefore having energy independent of the $t_1/t_2$ ratio [top red dashed curve in Fig.~5(d)].

The $t_1 t_2 t_2 t_1$ chain with PBC is a ${\cal SSC}(2)$ chain for $\theta_1=\arctan(t_1/t_2)$ and $\theta_2=\pi/2-\theta_1$,  and with a global hopping factor  $ t^{(2)} = \sqrt{t_1^2 + t_2^2}$.
This implies
\begin{eqnarray}
    \sin(\theta_1) &=& \dfrac{t_1}{\sqrt{t_1^2 + t_2^2}},\\
    \cos(\theta_1) &=& \dfrac{t_2}{\sqrt{t_1^2 + t_2^2}}.
\end{eqnarray}
The other parameters $t^{(n)}$, $n=1,0$, obtained by squaring this ${\cal SSC}(2)$ chain, are 
\begin{eqnarray}
    t^{(1)} &=& \sqrt{(\sin^2(\theta_1))^2 + (\cos^2(\theta_1))^2} =\dfrac{\sqrt{t_1^4+t_2^4}}{t_1^2+t_2^2},\\
    t^{(0)} &=& \sin^2(\theta_1) \cos^2(\theta_1)/(t^{(1)})^2 = \dfrac{t_1^2 t_2^2}{t_1^4+t_2^4},
\end{eqnarray}
and as one squares the original Hamiltonian, one obtains the following equations for the  spectra:
\begin{eqnarray}
    \varepsilon^{t_1 t_2 t_2 t_1 }(c) &=& t^{(2)} \varepsilon^{(2)}(c),\\
    (\varepsilon^{(2)}(c))^2  &=& t^{(1)} \varepsilon^{(1)}(c)+1,\\
    (\varepsilon^{(1)}(c))^2  &=& t^{(0)} \varepsilon^{(0)}(c)+1,\\
    \varepsilon^{(0)}(c) &=& c+1/c,
\end{eqnarray}
where $\varepsilon^{(0)}(c)$ is the spectrum of the uniform chain with unitary hopping terms (note that, in Ref.~\cite{Dias2021}, the uniform chain corresponds instead to ${\cal SSC}(0)$ with $\theta =\pi/4$ and $t^{(0)}$ differs from the above definition by a factor of $\sqrt{2}$).
The above relations can also be applied to the case of the $t_1 t_2 t_2 t_1$ chain connected to a ring of six sites, shown in Fig.~5(b). 
This spectrum in units of $t^{(2)}$ becomes the second one shown in Fig.~\ref{fig:squaringOBC}(a), $\varepsilon^{(2)}$. Squaring the respective Hamiltonian matrix, subtracting the identity matrix, and dividing by $t^{(1)}$, one obtains the third spectrum shown in Fig.~\ref{fig:squaringOBC}(a), $\varepsilon^{(1)}$. Repeating the latter step, now dividing by $t^{(0)}$ leads to the last spectrum shown in Fig.~\ref{fig:squaringOBC}(a), which is the spectrum of the uniform chain with a specific energy-dependent edge potential, $\varepsilon^{(0)}$. This edge potential generates the impurity-like curves shown in this spectrum. The edge state curves in the other spectra of Fig.~\ref{fig:squaringOBC}(a) are obtained from the impurity-like curves in the uniform chain spectrum, applying the relations in (10)-(13) in the opposite order, \textit{i.e.}, from bottom to top.

We now find the specific energy-dependent edge potential that determines the impurity curves in the uniform chain spectrum.
We start by determining the effective edge potential due to the ring. As before, we adopt $ t_2$ as the unit of energy, and we set $t_3=t_2$.
Starting from the leftmost site of the $t_1 t_2 t_2 t_1$ chain, the Lanczos method naturally generates bonding and anti-bonding states of the ring sites relative to the middle horizontal axis, and the tridiagonal Lanczos tight-binding matrix of the ring is
\begin{equation}
H_{\text Lanczos} =t_2 \left(
\begin{array}{cccccc}
 0 & 1 & 0 & 0 & 0 & 0 \\
 1 & 0 & 0 & 0 & 0 & 0 \\
 0 & 0 & 0 & \sqrt{2} & 0 & 0 \\
 0 & 0 & \sqrt{2} & 0 & 1 & 0 \\
 0 & 0 & 0 & 1 & 0 & \sqrt{2} \\
 0 & 0 & 0 & 0 & \sqrt{2} & 0 \\
\end{array}
\right).
\end{equation}
The anti-bonding states are decoupled from the $t_1 t_2 t_2 t_1$ chain and constitute a two-site cluster with a unitary hopping term, therefore leading to two levels of energy $\varepsilon=\pm t_3$, independent of the parameters of the chain. Consequently, in the decimation step, one only has to consider three hopping terms $\{t_{\bar{3}},t_{\bar{4}},t_{\bar{5}}\}=t_2 \{\sqrt{2}, 1, \sqrt{2}\}$ and onsite energies $\{ V_{\bar{3}}, V_{\bar{4}}, V_{\bar{5}}, V_{\bar{6}} \}=\{0,0,0,0\}$.
Decimation then leads to an effective potential 
$$
V^{(\bar{6})}_{\text{eff}}(c)= \dfrac{2 t_2^2}{\varepsilon(c)- \dfrac{t_2^2}{\varepsilon(c) - \dfrac{2 t_2^2}{\varepsilon(c)}} }.
$$
After this decimation step, one has an effective potential $V_{\text{eff}}^{(\bar{6})}(c)$ at the first site of $t_1 t_2 t_2 t_1$ chain. As done in the previous subsection, we replace this effective potential with a virtual site $\alpha$, and we have the relation
\begin{equation}
 V^{(\bar{6})}_{\text{eff}}(c)=\frac{t_{1 \rightarrow \alpha} t_{\alpha \rightarrow 1} }{\varepsilon (c)}=\frac{t_{ \alpha} ^2 }{\varepsilon (c)}.  
\end{equation}
Squaring the $t_1 t_2 t_2 t_1$  chain with the virtual site, the subchain with the virtual site is an SSH chain with staggered hopping parameters $t_ 2^2$ and $t_1^2$ and onsite energy $t_1^2+t_2^2$, except the first hopping term and onsite potential, which are respectively  $t_{\alpha} t_1$ and  $t_{ \alpha}^2 $, [see \ref{fig:squaringOBC}(b)]. The other subchain has uniform hopping terms $t_1 t_ 2$ and staggered local potentials $2 t_2^2$ and  $2 t_1^2$, except at the first site where the onsite potential is  $t_{\alpha}^2+t_1^2$. If we choose to square instead the  ${\cal SSC}(2)$ chain (obtained after dividing the $t_1 t_2 t_2 t_1$  chain with the effective edge potential by the global hopping factor  $ t^{(2)} = \sqrt{t_1^2 + t_2^2}$), then the two subchains parameters are divided by  $ (t^{(2)})^2$. This sequence of steps is shown in Fig.~\ref{fig:squaringOBC}(b).

In the following steps, one can choose either of the subchains since they have the same spectrum \cite{Ezawa2020,Marques2021a} (the SSH-like subchain may have an extra zero state due to the extra virtual site, but this level falls into a dispersive band). If the latter is chosen, one has to apply the step shown in Fig.~\ref{fig:virtualsite}(b), which is slightly more complicated.
Therefore, we choose the SSH-like chain to proceed and obtain the spectrum of the uniform chain as shown in Fig.~\ref{fig:squaringOBC}(a).  
The following steps repeat the ones shown in Fig.~\ref{fig:virtualsite}(a), now starting from the SSH chain with a virtual site.
Instead of repeating them, we reduce this SSH chain to the case of the previous subsection by decimating the virtual site to obtain an effective potential at the first real site of the SSH subchain. Subtracting the global onsite energy $t_1^2+t_2^2$, this effective potential becomes 
\begin{equation}
 V^{SSH}_{\text{eff}}(c)=\frac{(t_{\alpha} t_{1} )^2}{\tilde{\varepsilon }(c)-t_{\alpha}^2}.  
\end{equation}
where $\tilde{\varepsilon }(c)$ is the energy of the SSH chain before the energy shift, that is, $\tilde{\tilde{\varepsilon }}(c)=\varepsilon ^{SSH}(c)=\tilde{\varepsilon }(c) -t_1^2+t_2^2= ( t^{(2)} )^2  t^{(1)} \varepsilon^{{\cal SSC}(1)}(c) $.
The expression for $c$ in (\ref{eq:cformula}) now becomes
\begin{equation}
  c=  \dfrac{t_1^2 t_2^2}{V^{SSH}_{\text{eff}}(c)\tilde{\tilde{\varepsilon }}(c)-t_1^4}
\end{equation}
where we have used the fact that the hopping terms in the SSH chain are  $t_ 2^2$ and $t_1^2$.
Using the relation between $\tilde{\tilde{\varepsilon }}(c)$ and $\varepsilon^{(0)}(c)$, the solutions of this equation generate the colored curves in the spectrum of the uniform chain in Fig.~\ref{fig:squaringOBC}(a). Using (10)-(12), the colored curves in the other spectra of  Fig.~\ref{fig:squaringOBC}(a) are obtained. All of these colored curves reproduce exactly the numerical edge state curves obtained with the full diagonalization of the Hamiltonian.

Note that the virtual site at the SSH subchain may add a spurious solution of energy $\tilde{\tilde{\varepsilon }}(c)=0$. These spurious solutions may occur whenever we add a virtual site during the squaring process. Their energies correspond to the folding levels (see App.~D) associated with the chiral symmetries that appear at each step of the squaring process in the pristine ${\cal SSC}(n)$ spectrum \cite{Dias2021,Marques2021}.

\section{Conclusion}
This paper presented a complex band structure analysis of one-dimensional square and high-root topological insulators. We have shown that the edge state bands observed in HRTIs correspond to sections of impurity bands of uniform tight-binding chains. These edge-state bands can be probed by introducing clusters at the boundaries of the TI chains, and a simplified topological characterization of HRTIs was proposed for this more general situation that relies on two key factors: the existence of these edge-state bands in the infinite system and the restrictions imposed by the boundary conditions. Closed-form equations were obtained to determine the edge state levels without the need for the diagonalization of real space or bulk Hamiltonians. These equations reflect the mapping of the edge state levels of  HRTIs onto impurity Tamm states of the uniform chain with edge energy-dependent potentials.
This analysis will be extended to two-dimensional TIs (topological ribbons, weak TIs, higher-order TIs, etc.) in a future paper.

In the systems studied here, the usual topological characterization using topological invariants extracted from the Bloch bands is insufficient to characterize the multiple regions in the phase diagram induced by the generalized boundary conditions, and, therefore, while briefly commented in App.~C, it was not explored in the main text. More precisely, the usual invariants (and the bulk-edge correspondence) characterize the behavior of topological insulators for a particular simple boundary choice, that is, open boundary conditions.

The edge-state bands of 1D HRTIs, like Bloch bands, can be experimentally probed by using artificial platforms such as acoustic lattices \cite{Yan2020,Cheng2022,Wu2023} or photonic lattices \cite{Cui2023,Kang2023,Yan2021}. This can be achieved by adding a cluster at the edges of the HRTI chain, leading to the emergence of multiple edge states in a particular edge-state band as discussed in Sec.~V, or by examining the energy and form of a single edge state as it moves along the edge-state band curve while an edge potential is continuously varied.

\section*{Acknowledgments}

The authors thank J. Rossier for important discussions. R.G.D., L.M., and A.M.M. developed their work within the scope of the Portuguese Institute for Nanostructures, Nanomodelling, and Nanofabrication (i3N) Projects No. UIDB/50025/2020, No. UIDP/50025/2020, and No. LA/P/0037/2020, financed by national funds through the Funda\c{c}\~ao para a Ci\^encia e Tecnologia (FCT) and the Minist\'erio da Educa\c{c}\~ao e Ci\^encia
(MEC) of Portugal. 
L.M. acknowledges the financial support from FCT (grant No. PTDC/FIS-MAC/2085/2020).
L.M. acknowledges support from the European Union (Grant FUNLAYERS-101079184).
A.M.M. acknowledges financial support from FCT through the work Contract No.~CDL-CTTRI-91-SGRH/2024.

\section*{Data availability}
The data that support the findings of this article are available upon reasonable request from the authors.

\appendix

\section{Infinite uniform tight-binding chain}
In the Sec.~\ref{sec:intro}, we mentioned that the set of eigenvectors (and respective eigenvalues) of the thermodynamic limit of the eigensolutions of the finite tight-binding chain is included as a subset of the eigensolutions of the infinite chain tight-binding Hamiltonian. In fact, the eigenvalue equations of the infinite chain even admit complex eigenvalues. The eigenvalue equation of an infinite uniform tight-binding chain (with hopping parameter $t=1$) is
\begin{equation}
\epsilon \psi_j = \psi_{j+1}+\psi_{j-1},
\end{equation}
with $j$  integer.  If we assume the ansatz $\psi_j=z^j$,  where $z$ is a complex number, one obtains
\begin{equation}
\epsilon(z)=z+1/z=\vert  z \vert e^{i \theta} + \dfrac{1}{\vert  z \vert } e^{-i \theta}.
\end{equation}
The condition of real energies implies $\vert z \vert$=1 (Bloch states) or $\theta=0,\pi$ (impurity/edge-like states).
So, in the case of  Bloch states, the usual energy dispersion is obtained (see Fig.~\ref{fig:1}), 
\begin{equation}
\epsilon(k)=e^{i k}+e^{-i k}.
\end{equation}
In the case of  edge states, the  energy dispersion becomes 
\begin{equation}
	\epsilon(c)=c+1/c ,
\end{equation}
with $c$ real. The energy dispersion of Bloch and edge-like states in the infinite chain is shown in Fig.~1. The absolute value of $c$ determines whether one has a left decaying edge-like state ($\vert c\vert<1$) or a right decaying one ($\vert c\vert>1$).

\section{Finite or semi-infinite uniform tight-binding chain}
At this point, the attentive reader wonders if the number of Bloch and edge state solutions of the eigenvalue equation does not exceed the dimensionality of the Hilbert space and why the edge states are not orthogonal when they have different energies. To explain these two points, we must now consider the introduction of boundary conditions, that is, a finite-size (or semi-infinite)  tight-binding chain.

The eigenstates of a uniform tight-binding open chain with N sites are harmonic states $\psi_j=\sqrt{\frac{2}{N+1}} \sin (k.j)$ with $k=\frac{\pi}{N+1}.n$, and $n=1,\cdots,N$. The states have zero amplitudes at the virtual sites $j=0$ and $j=N+1$ and can be viewed as the anti-symmetric combination of the Bloch states $\vert k \rangle$ and  $\vert -k \rangle$ of the infinite chain. Such a combination has zero amplitudes at sites $j=q(N+1)$ with $q$ an integer, and this implies that the sequence of amplitudes between two consecutive zeros will be an eigenstate of the $N$-site finite chain with the same energy as that of the infinite chain. The number of harmonic states equals $N$, and no edge state is present.

One can force the existence of an edge state in the finite uniform chain by adding a local potential at an edge site (this implies one less harmonic state since the total number of states is equal to the dimension of the Hilbert space). Assuming a local potential (energy independent for now) at the left edge site (see third diagram of Fig.~\ref{fig:stepsdiagram}), and a semi-infinite chain (so that only a left decaying component is present), the edge state energy and the respective $c$-value can be determined in two different but equivalent ways.

In a first method, one assumes the ansatz $\psi_j=c^j$ to be valid in the bulk region,  obtaining the relation between the energy and $c$ from the eigenvalue equation $\epsilon \psi_j = \psi_{j+1}+\psi_{j-1}$. This leads to the energy dispersion $\epsilon(c)$ of the infinite chain given above. 
Second, one addresses the eigenvalue equation at the first site,
\begin{equation}
\epsilon(c)  \psi_1 = \psi_2+V_{\text{eff}} \psi_1
\label{eq:edgepot}
\end{equation}
and by replacing the ansatz amplitudes (which are determined by the bulk region eigenvalue equations), one gets 
\begin{equation}
\epsilon(c) = c + V_{\text{eff}},
\end{equation}
and therefore 
\begin{equation}
 c =\dfrac{1}{V_{\text{eff}}} .
\end{equation}
This last result implies the left edge state appears when $\vert V_{\text{eff}} \vert >1$ and evolves from the $k=0$ ( $k=\pi$) state when $V_{\text{eff}}>0$ ($V_{\text{eff}}<0$)  as one can also conclude from the energy plots in  Fig.~1. 

The second method  reflects the assumption that the edge state of the semi-infinite chain with a local potential at the left edge site (again, see third diagram of Fig.~\ref{fig:stepsdiagram}) is a truncated edge state of the infinite chain and, therefore, (\ref{eq:edgepot}) should be equivalent to 
\begin{equation}
	\epsilon(c)  \psi_1 = \psi_{2}+ \psi_{0},
	\label{eq:edgepot2}
\end{equation}
that is, 
\begin{equation}
	V_{\text{eff}} \psi_{1}=  \psi_{0}.
\end{equation}
This leads to the same relation between $c$ and $V_{\text{eff}}$. Note that while being equivalent approaches, the second method does not require knowledge of the edge state energy dispersion. 

\section{Infinite  SSH chain}
The same reasoning of the previous section can be followed in the case of the infinite SSH chain.
Again,  we assume the ansatz 
\begin{equation}
\ket{\psi_{edge}(z)}_{j}=z^{j}\ket{u(z)}=z^{j}\begin{bmatrix}\psi_{A}(z)\\
\psi_{B}(z)
\end{bmatrix},
\end{equation}
leading to the ``bulk'' Hamiltonian of the infinite SSH chain
 \begin{equation}
H(z)=\begin{bmatrix}0 & t_{1}+t_{2}/z\\
t_{1}+t_{2}z & 0
\end{bmatrix},
\end{equation}
and its energy dispersion can be obtained from
 \begin{equation}
\varepsilon(z)^2=(t_{1}+t_{2}/z)(t_{1}+t_{2}z).
\end{equation}
Imposing again the condition of a real spectrum, we find  $\vert z \vert$=1 (Bloch states) or $z$ real, but confined to the intervals $[-t_2/t_1,-t_1/t_2]$ and $]0,\infty]$ (where we have assumed $t_1$ and $t_2$ positive and $t_2>t_1$). In this case, we have
 \begin{equation}
\varepsilon(z)=\pm\sqrt{(t_{1}+t_{2}/z)(t_{1}+t_{2}z)},
\end{equation}
with eigenstates
 \begin{equation}
\vert u(z)\rangle=\frac{1}{\sqrt{2}}\begin{bmatrix}1\\
{\displaystyle \mp\sqrt{\frac{t_{1}+t_{2}z}{t_{1}+t_{2}/z}}}
\end{bmatrix}.
 \end{equation}
 
The Bloch and edge state Hamiltonian, $H(k)$ and $H(c)$, the respective energy dispersions $\varepsilon(k)$ and $\varepsilon(c)$, and respective eigenstates $\vert u(k)\rangle$ and  $\vert u(c)\rangle$ of the infinite SSH chain are obtained from  $ z \rightarrow e^{ik}$ and  $z \rightarrow c$.

The last expressions, being common to  Bloch and edge-like eigenstates of the infinite chain, allow us to introduce a generalized topological characterization of the SSH chain where the behavior of the Bloch and edge-like bands are analyzed together. In Fig.~\ref{fig:1}, we show in (c2) edge-state and Bloch bands and in (b2) the regions of the $z$ -plane corresponding to the real spectrum of the infinite SSH chain with $t_1=0.5 $ and $t_2=1$. The edge state and Bloch bands share the states at the inversion-invariant momenta of the Bloch bands. These shared points are indicated as the same-colored circles in the three plots. 

The topological characterization of the SSH model usually relies on the calculation of Zak's phase of the Bloch bands \cite{Zak1989}, and the topological phase transition is signaled by the discontinuous change of the Zak's phase from 0 to $\pi$ or vice-versa \cite{Zak1989}. This change in Zak's phase reflects the exchange of the bottom band's top state and the top band's bottom state at the topological transition \cite{Asboth2016,Zak1985,Fang2012}. These states are eigenstates of the inversion operator with different parity values, and the Zak's phase can be rewritten as the phase of the product of each Bloch band's top and bottom states. It is easy to conclude that equally \emph{the topological transition is signaled by the exchange of the parity values of the top and bottom states of the central edge-state band} (but no exchange occurs on the top and bottom impurity-state bands).
Note that, at the transition point, the edge band reduces to the degenerate states at $c=\pm 1$ ($k=0,\pi$), which are also common to the bulk bands.

As a final comment in this section, we would like to point out that 
\begin{equation}
f(z)={\displaystyle\frac{t_{1}+t_{2}z}{t_{1}+t_{2}/z}}
\end{equation}
can be interpreted as a conformal mapping that distinguishes between the topological trivial and non-trivial phases. This particular function pops up in several topological characterizations of the SSH chain. Besides Zak's phase (and winding number), another example is the particle density calculation at the first site of the SSH chain with OBC generated when the bottom band is filled \cite{Marques2019}.

\section{Squarable Sine-Cosine Chains}
Here, we review the construction method \cite{Dias2021} of the bipartite Squarable Sine-Cosine chains, ${\cal SSC}(n)$, which are the simplest strictly 1D HRTIs.  

Sine-Cosine models  [$\mathcal{SC}(n)$] are a particular set of generalized Su-Schrieffer-Heeger models with $2n$ sites in the unit cell [SSH($2n$)], that, when squared, generate a block-diagonal matrix with one of the blocks corresponding to a subchain with unitary uniform local potentials. 
The $\mathcal{SC}(n)$-chain is defined by hopping terms between adjacent sites with parameters:
\[
h_{j} = t^{(n)} \sin \theta_j^{(n)}, \quad w_{j} = t^{(n)} \cos \theta_j^{(n)},
\]
where $t^{(n)}$ is the global hopping factor, and $\{\theta_j^{(n)}\}_{j=1}^{2^n}$ are the angles determining the hopping terms sequence.

Squarable Sine-Cosine chains, ${\cal SSC}(n)$, are a subset of Sine-Cosine chains that, when squared, always generate a block that is again a Sine-Cosine model if an energy shift is applied and the energy unit is normalized.
For $\mathcal{SSC}(n)$ chains with PBC, the angles $\{\theta_j^{(n)}\}_{j=1}^{2^n}$ are recursively defined from $\mathcal{SSC}(n-1)$. Specifically:
\begin{eqnarray}
	t^{(n-1)} \sin \theta_{j}^{(n-1)}&=& \cos \theta_{2j-1}^{(n)}  \sin \theta_{2j}^{(n)} \label{eq:1}\\
t^{(n-1)} \cos \theta_{j}^{(n-1)}&=& \cos \theta_{2j}^{(n)}  \sin \theta_{2j+1}^{(n)} 
\label{eq:2}
\end{eqnarray}
for $j=1, \ldots, 2^{n-2}$ with the periodic condition $2^{n-1} + 1 \equiv 1$. The global hopping factor in the  SSC$(n-1)$ chain is given by
\begin{equation}
	t^{(n-1)}= \sqrt{( \cos \theta_{2j-1}^{(n)}  \sin \theta_{2j}^{(n)})^2+(\cos \theta_{2j}^{(n)}  \sin \theta_{2j+1}^{(n)} )^2}
\end{equation}
for any value of  $j$.  
These equations can be iteratively solved for each $n$, building up from $\mathcal{SSC}(0)$, and the knowledge of the global hopping factors at each step, $t^{(n)}$,   is enough to determine the band structure of the $\mathcal{SSC}(n)$ chain (the energy shifts are always 1).

Chiral symmetry is present in all  Sine-Cosine chains and the zero-energy edge states of a SSC$(j)$ chain (with $j=1,\dots,n$)  with open boundary conditions become finite energy edge states in non-central band gaps of the $\mathcal{SSC}(n)$  chain with energies \cite{Dias2021}
\begin{align*}	
&\pm 1,\\
&\pm \sqrt{1\pm t^{(n-1)}},\\
&\pm \sqrt{1\pm t^{(n-1)}\sqrt{1\pm t^{(n-2)}}},\\
&\vdots,\\
&	\pm \sqrt{1\pm t^{(n-1)}\sqrt{1\pm t^{(n-2)}\sqrt{\ddots \sqrt{1\pm t^{(1)}}}}}.
\end{align*}
\section{Cluster diagonalization approach}
In Sec.~III, a Lanczos tridiagonalization of the cluster sites was described as an intermediate step to find the effective edge potential of the TI chain. An alternative method would be to diagonalize the cluster Hamiltonian \cite{Marques2020}, $H^{\text cluster} \vert \psi^{\text cluster}_i \rangle = \varepsilon^{\text cluster}_i \vert \psi^{\text cluster}_i \rangle$, so that the effective potential at the edge of the TI chain would now be 
\begin{equation}
 V^{TI}_{\text{eff}}(c)=\sum_i  \frac{\vert  t_{i} \vert^2}{\varepsilon (c)-\varepsilon^{\text cluster}_i }.  
\end{equation}
where $t_i=\langle  \psi^{\text cluster}_i \vert H \vert 1, \text{TI chain} \rangle $.
For example, in the case of the $t_1 t_2 t_2 t_1$ chain connected to a ring of six sites, shown in Fig.~5(c), $H^{\text cluster} $ is the Hamiltonian of a linear chain with five sites, hopping term $t_3$, and open boundary conditions. The respective energies are 
   $$
  \varepsilon^{\text cluster}_m \;=2t_3 \cos\!\left(\frac{\pi m}{6}\right),\quad m = 1, 2, \ldots, 5.
   $$
and the eigenstates are 
$$
   |\psi^{\text cluster}_m\rangle \;=\; \sqrt{\frac{1}{3}}\,\sum_{n=1}^{5} 
   \sin\!\left(\frac{\pi \, m\,n}{6}\right)\,|n\rangle.
   $$
with $n = 1, 2, \ldots, 5$ labeling the lattice sites of the cluster chain in a clockwise fashion. Therefore 
\begin{eqnarray}
    t_m &=& \langle  \psi^{\text cluster}_m \vert H \vert 1, \text{TI chain} \rangle \nonumber \\
&=& 
\sqrt{\frac{1}{3}}\left[
   \sin\!\left( \frac{\pi \, m}{6} \right) + \sin\!\left( \frac{\pi \, 5\,m }{6} \right)\right]\,t_3.
   \nonumber
\end{eqnarray}

and
\begin{equation}
 V^{TI}_{\text{eff}}(c)=\sum_m  \frac{   \frac{1}{3}\vert  \left[
   \sin\!\left( \frac{\pi \, m}{6} \right) + \sin\!\left( \frac{\pi \, 5\,m }{6} \right)\right]\,t_3 \vert^2}{\varepsilon (c)-t_3 \cos\!\left(\frac{\pi m}{N+1}\right) }.  
\end{equation}

\bibliography{bibliografia2}

\end{document}